\documentstyle[twocolumn,aps,epsf]{revtex}
\begin{document}
\twocolumn[\hsize\textwidth\columnwidth\hsize\csname
@twocolumnfalse\endcsname
\title{
Symmetry of `molecular' configurations of interacting electrons in a
quantum dot in strong magnetic fields
}
\author{Hiroshi Imamura, Peter A. Maksym$^{(a)}$ and Hideo Aoki
\\
{\small Department of Physics, University of Tokyo, 
Hongo, Tokyo 113, Japan \\ 
$(a)$ Department of Physics and Astronomy,
University of Leicester, \\ Leicester LE1 7RH, United Kingdom}
}
\maketitle
\begin{abstract}
A molecular description for magic-number configurations of 
interacting electrons in a quantum dot in high magnetic fields 
developed by one of the 
authors has been elaborated for four, five and six electron dots.  
For four electrons, the magic spin-singlet states are found to 
alternate between two different resonating valence bond (RVB)-like states.
For the five-electron spin-polarized case, the molecular description is 
shown to work for the known phenomenon of magic-number 
sequences that correspond to both the $N$-fold symmetric ring 
configuration and a $(N-1)$-fold symmetric one with a 
center electron. 
A six-electron dot is shown here to have an additional feature in which 
inclusion of quantum mechanical mixing 
between classical configurations, which are deformed and degenerate, 
restores the $N$-fold symmetry and reproduces the ground-state 
energy accurately.
\end{abstract}

\vspace*{0.12cm}

\vspace*{-0.50cm}
\vskip2pc]
\section{introduction}
Quantum dots are two-dimensional (2D) electron systems confined in a 
small ($\sim$ few
hundred nanometer) region and have attracted much attention as 
artificial atoms~\cite{kastner}\cite{tarucha}.  
A remarkable feature of these `atoms' is their controllability.  
A dramatic feature in fact arises when quantum dots 
are placed in high magnetic fields, where specific 
total-angular momentum ($L$) states are stabilized 
with downward
cusps in the energy against $L$.  
The cusps, originally found from 
numerical calculations \cite{girvin}\cite{map-prl}, are 
called `magic numbers' in the angular momenta.  
When the magnitude of magnetic field is varied, 
the true ground state changes from one magic-number state to another.  
Some of the cusps occur at the same fractional filling factors as 
the fractional quantum Hall effect in bulk systems. 

In a different light, quantum dots may be regarded as a
2D `molecule' that consists only of electrons, as contrasted 
with the ordinary atoms where electrons are bound to nuclei.  
The quantum 
states in strong magnetic fields can in fact 
be treated with a molecular description, which has been 
developed by one of the present authors \cite{map-eckart}. 
In this approach we start from 
the classical equilibrium configuration and make an 
expansion of the Hamiltonian 
to derive a quantum mechanical 
rotational-vibrational Hamiltonian to obtain 
anti-symmetrized wavefunctions.  
In a previous work~\cite{map-eckart}
this approach has been tested for 
ground and excited states of three electrons, and 
was found to give energies as accurate as within 0.1\% 
of the exact-diagonalization result for
$L \sim 30$ in a parabolic confining
potential. 

Here we extend the theory to four, five, and six electrons 
in search of features for larger number of electrons. 
All our results are for a GaAs dot with a typical 
experimental situation of 4meV confinement energy in a 
magnetic field of 20T.
\section{Magic numbers}
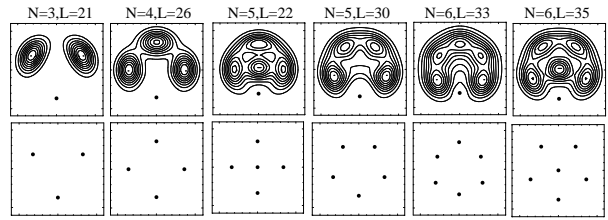
\begin{figure}
    \epsfxsize=8cm
\centerline{\hbox{
      \epsffile{eq-pc.epsi}  }}
  \vspace{0.5cm}
  \caption{Quantum mechanical density 
  correlation with one electron fixed at the position indicated 
  by a dot (upper panels) as compared with classical equilibrium
    point-charge configurations (lower panels) 
    for typical magic-number 
states. For $N=5, 6$, $N$-fold and $(N-1)$-fold symmetric 
states occur and give rise to two sequences of 
magic numbers with intervals $N$ and $N-1$, respectively.
    }
  \label{fig:eq-pc}
\end{figure}

The magic numbers are known to 
depend on the number of electrons, $N$, and total spin, $S$. 
For fully spin-polarized electrons, 
magic $L$'s are 
3,6,9,12,... for $N=3$ and $L=6,10,14,18,...$ for $N=4$. 
Namely, the magic numbers occur with an interval of $N$, 
or more precisely $L=
N(N-1)/2+N\times{\rm integer}$.  
Another sequence of magic numbers, having an 
interval $N-1$ with $L= N(N-1)/2+(N-1)\times{\rm integer}$, 
\cite{map-eckart}\cite{ruan}\cite{kuramoto} appears for 
six spin-polarized electrons. For five spin-polarized electrons, the main 
sequence has an interval $N$ and the ground state for 
any the magnetic field belongs 
to the sequence. 
If one looks more closely at the $L$- dependence, 
however, there are 
weak cusps with an interval $N-1$, and we can 
call these states excited magic-number states.  
It is known that 
an $N$-fold ring has the magic number states
with interval $N$, while an $N-1$-fold ring with a particle at the center 
has an interval $N-1$.  
If we look at the quantum mechanical 
density correlation function 
in Fig.~\ref{fig:eq-pc}, where we 
fix the position of one electron to plot the density of others, 
we can indeed identify both $N$- and $N-1$-fold rings.  

\section{Eckart-frame theory}
The molecular description provides not only a way of understanding the 
magic numbers, but also a quantitatively accurate 
description of the ground-state 
energy in the strong field regime \cite{map-eckart}.
The starting point is the remarkable correspondence between 
the quantum mechanical configurations 
(upper panels of Fig.1) and the classical 
equilibrium configurations 
(lower panels) of point charges interacting with the Coulomb repulsion.  
The physical picture of the system in the high field
regime is then that of an `electron molecule' which rotates with 
total angular momentum $L$ and vibrates around its classical
equilibrium configuration. 
We can first harmonically expand the Hamiltonian around the 
classical configuration.  
The vibration is treated in a similar 
way to that of chemical molecules, 
and this is facilitated by the use of a molecular-fixed reference 
frame called the Eckart frame in which the rotational and vibrational motions
decouple to first order \cite{map-eckart}.  
We then diagonalize the
Hessian matrix for the harmonically expanded Hamiltonian 
and obtain the effective potential.  Using the
effective potential and taking account of Coriolis coupling, we 
find the rotational-vibrational states.  The final important 
ingredient is the anti-symmetrization of the wavefunction.

When the classical equilibrium configuration has $C_m$ symmetry, 
a group-theoretical consideration dictates that 
antisymmetrization enforces a wavefunction to have a factor 
\begin{equation}
  \sum_{t=0}^{m-1} (\pm 1)^{t} \! \exp \!
        \left[
        \frac{-2\pi i t}{m}
                \left(
                L_{_{RM}} \! + k_{s} \! + \!\!\sum_{i=1}^{2N-3} n_{i}k_{v}(i)
                \right)
        \right],
\end{equation}
where $n_{i}, k_{v}(i)$ are the quantum number and phase parameter
corresponding to the $i$\/th normal mode, respectively, 
while $k_{s}$ is the phase
parameter for the spin wavefunction. It follows that
the wavefunction vanishes unless
\begin{equation}
L_{_{RM}} + k_{s} + \sum_{i=1}^{2N-3} n_{i}k_{v}(i)
\equiv
\left\{
\begin{array}{lll}
0 & {\rm mod} \ m,& m~{\rm odd}\\
m/2 &  {\rm mod} \ m,& m~{\rm even}
\end{array}
\right.
\label{eq:magic-sequence}
\end{equation}
This criterion reproduces the correlation between the 
$N$- and $(N-1)$-fold rings and the interval of magic numbers.

After quantum numbers that satisfy this condition have been found, 
the total energy can be written in the form
\begin{equation}
E=E_{_{CM}}+E_{cl}+E_{zp} +E_{x}(n_{1},\cdots,n_{_{2N-3}})
\end{equation}
where $E_{_{CM}}$ is the energy of center-of-mass motion, $E_{cl}$ is
that of the classical configuration, $E_{zp}$ is the quantum zero-point
energy and $E_{x}(n_{1},\cdots,n_{_{2N-3}})$ is the excitation
energy of the vibrational state.
\section{Four electrons}
\begin{figure}
    \epsfxsize=8cm
\centerline{\hbox{
      \epsffile{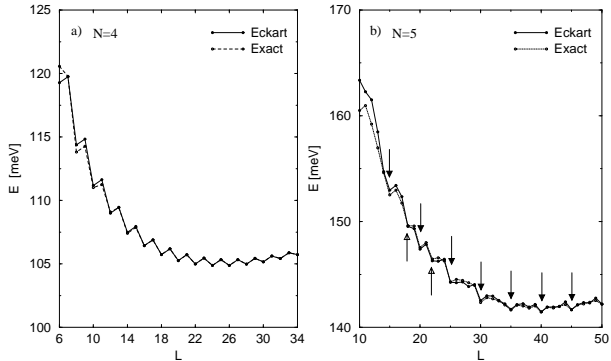}  }}
  \vspace{0.5cm}
  \caption{
    (a) Lowest energies against the total angular momentum 
    $L$ for $N=4$, $S=0$.\newline
    (b) The same for $N=5$, $S=5/2$.
    Solid (open) arrows indicate the magic numbers with interval $N$ ($N-1$).
    In each figure, the solid line with filled circles represents the
    result of Eckart-frame theory while the dotted line with open
    circles gives the exact diagonalization result.
    }
  \label{fig:ge-n4-n5}
\end{figure}

\begin{figure}
    \epsfxsize=8cm
\centerline{\hbox{
      \epsffile{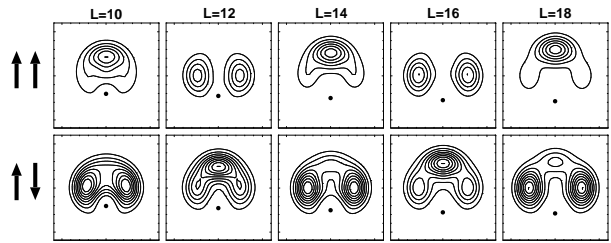}  }}
  \vspace{0.5cm}
  \caption{
    Spin-resolved density correlation functions for 
    $N=4$, $S=0$ magic-number states.  We
    fix an up-spin at the position indicated with a filled circle 
    to contour-plot the density of parallel spins (upper panels) 
    and anti-parallel spins (lower panels).  
    }
  \label{fig:pc-n4-s0-bw}
\end{figure}

We have calculated energies and density correlation functions for ground states
against total angular momentum for all the spin configurations
using both exact diagonalization and Eckart-frame theory.
The energies for spin-singlet ($S=0$) states 
are plotted in
figure~\ref{fig:ge-n4-n5}, which 
shows that the Eckart-frame theory produces accurate
ground state energies.  We have also checked that ground-state energies
obtained by Eckart-frame theory for all the 
other spin configurations agree very well with those of the exact
diagonalization.  
One can readily see from equation ~\ref{eq:magic-sequence}
that the interval between magic numbers is affected by 
the total spin $S$.  For the spin-polarized case,
$S=2$ here, 
the magic numbers are $L=6, 10, 14, 18, ...$ and these are explained
by equation~\ref{eq:magic-sequence} with $k_s=0$.
In the case of $S=0$ states
magic numbers are $L=6, 8, 10, 12, 14, 16, 18, ...$.  The 
spin-resolved 
correlation functions are displayed in fig.~\ref{fig:pc-n4-s0-bw}, 
which show
that there are two different sequences; one has $L=10, 14,...$
while the other has $L=12, 16,...$.  
In the Eckart-frame theory,
these magic numbers are characterized by different $k_{s}$.  The
former corresponds to $k_s=0$ and the latter to $k_{s}=2$.
Except for a normalization, the spin wavefunctions are written as
\begin{eqnarray}
k_s=0 &:& |\uparrow \uparrow \downarrow \downarrow \rangle
       + |\downarrow \uparrow \uparrow \downarrow \rangle
       + |\downarrow \downarrow \uparrow \uparrow \rangle
       + |\uparrow \downarrow \downarrow \uparrow \rangle \nonumber\\
      &&\mbox{\hspace{0cm}}-2[|\uparrow \downarrow \uparrow \downarrow \rangle
          +|\downarrow \uparrow \downarrow \uparrow \rangle ],\label{eq:ks0}
\\
k_s=2 &:& |\uparrow \uparrow \downarrow \downarrow \rangle
       - |\downarrow \uparrow \uparrow \downarrow \rangle
       + |\downarrow \downarrow \uparrow \uparrow \rangle
       - |\uparrow \downarrow \downarrow \uparrow \rangle 
.
\label{eq:ks2}
\end{eqnarray}
We can rewrite these in terms of spin-singlet pairs as
\unitlength 1mm
\begin{eqnarray}
&&
\mbox{
\begin{picture}(50,10)
\put(0,2){$k_s = 0:$}
\put(15,0){\framebox(1,6){}}
\put(20,0){\framebox(0.9,6){}}
\put(16,0){\line(20,0){4}}
\put(16,6.1){\line(20,0){4}}
\put(24.5,2){$-$}
\put(30,0){\framebox(6,1){}}
\put(30,5){\framebox(6,1){}}
\put(30,0){\line(0,1){6}}
\put(36.1,0){\line(0,1){6}}
\end{picture}
}\\
&&
\mbox{
\begin{picture}(50,10)
\put(0,2){$k_s = 2:$}
\put(15,0){\framebox(1,6){}}
\put(20,0){\framebox(0.9,6){}}
\put(16,0){\line(20,0){4}}
\put(16,6.1){\line(20,0){4}}
\put(24.5,2){$+$}
\put(30,0){\framebox(6,1){}}
\put(30,5){\framebox(6,1){}}
\put(30,0){\line(0,1){6}}
\put(36.1,0){\line(0,1){6}}
\put(38,2){$\equiv$}
\put(42,0){\framebox(6,6){}}
\put(42,0.7){\line(1,1){5.3}}
\put(42.7,0){\line(1,1){5.3}}
\put(42,5.3){\line(1,-1){5.3}}
\put(42.7,6){\line(1,-1){5.3}}
\end{picture}
}
\end{eqnarray}
where 
\begin{picture}(8,10)
\put(0,0){\framebox(6,1){}}
\end{picture}
$\equiv \frac{1}{\sqrt{2}}(|\uparrow\downarrow \rangle - 
|\downarrow\uparrow \rangle )$.
These wavefunctions are the same as the RVB (resonating valence bond)
states considered for the Hubbard model~\cite{aoki-hubbard}.
The configuration of the spin resolved density correlation in
Fig.~\ref{fig:pc-n4-s0-bw} is indeed explained from equations~\ref{eq:ks0},\ref{eq:ks2}.
\section{Five electrons}
In the case of five electrons, there are two different classical
equilibrium configurations: a pentagon and a square with one 
particle at the center. These configurations are shown in 
Fig.~\ref{fig:eq-pc}. We have calculated lowest-energy states against 
$L$ in the Eckart-frame theory for 
both of the configurations, and have plotted the 
lowest of these two energies in 
Fig.~\ref{fig:ge-n4-n5}.  
The difference from the exact-
diagonalization result 
is less than 0.2 \% for $L \sim 40$.

In Fig.~\ref{fig:ge-n4-n5}, we have indicated the magic numbers with 
the interval $N$ (solid-down-arrows) and the interval $N-1$ (open-up-arrows).  
If we look at in Fig.1 the density correlation for the states belonging to 
the $N$-sequence ($L=30$) or the $(N-1)$-sequence ($L=22$), 
the shape of the molecule does indeed 
possess the $N$-fold or $(N-1)$-fold symmetries, respectively.
\section{Six electrons}
For the six electron system, we have again two kinds of classical
equilibrium configurations: a six-fold ring and a pentagon 
with a center particle as 
shown in Fig.1.  If we more closely look at 
the six-fold ring, the classical configuration 
is distorted from the regular hexagon and there are doubly-degenerate 
three-fold equilibria, which would naively imply magic 
numbers with an interval of three.

\begin{figure}
    \epsfxsize=8cm
\centerline{\hbox{
      \epsffile{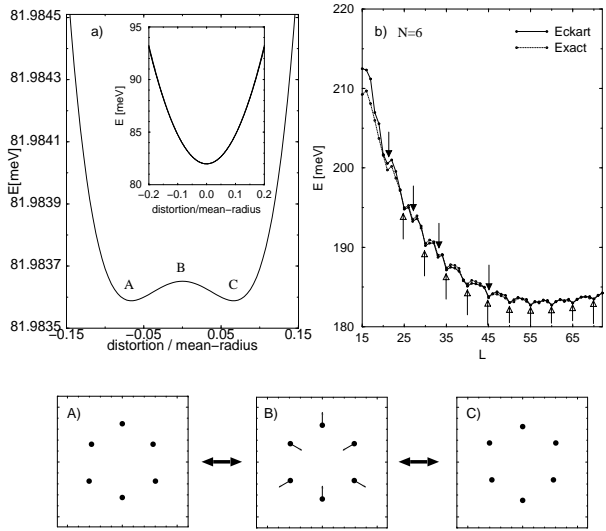}  }}
  \vspace{0.5cm}
  \caption{
    (a) The classical potential-energy surface 
    for $N=6$, $S=3$, $L=33$, plotted against the
    distortion in the lowest normal mode. 
    Double minima correspond to doubly-degenerate, deformed 
    classical equilibria indicated in insets A and C, 
    while the saddle point B is a
    regular hexagon.  In the inset of (a), the vibrational potential
    energy is plotted against the distortional vibration.\newline 
    (b) Lowest energies against $L$ for $N=6$, $S=3$. The solid line
    with filled 
    circles is obtained in the Eckart-frame theory with the 
    quantum mixing, while the dotted line 
    with open circles the exact diagonalization result.
    }
  \label{fig:dist6}
\end{figure}

The exact-diagonalization result for the 
magic numbers for six spin-polarized 
electrons, however, consists of a
series having period of six ($L=21,27,33$) along with another having
period of five ($L=25,30,35,40,45,...$). 
The period of six may 
at first seem inconsistent with the
three-fold classical configurations.  
However, this discrepancy can be resolved 
by noting that the six-fold symmetry is restored when a
quantum mechanical 
mixture of the rotational-vibrational states centered at the 
two three-fold configurations is used to construct the 
approximate electron state.  In Fig.~\ref{fig:dist6}, the classical
energy of a rotating configuration ($L=33$, no vibration) is plotted 
against the distortion (see Fig.~\ref{fig:dist6} B) that couples the
three-fold rings.
The saddle point is the regular hexagon while the two minima
are the rings distorted in staggered 
directions.  The height of the barrier (about 0.0001meV) is much 
smaller than the typical lowest quantum vibrational energy (about 1.0 meV),
so that they are almost completely mixed 
to result in a restored six-fold symmetry.

To calculate the energy of the hexagonal states 
we approximate the vibrational potential energy. A six-electron 
molecule has nine in-plane normal modes and the lowest has the same 
displacement pattern as distortion shown in Fig. \ref{fig:dist6}. 
The vibrational potential depends on $L$ and is almost parabolic for
$L \geq 15$, but for the normal mode shown in Fig. \ref{fig:dist6} 
there is some small extra structure which originates from the 
double minimum in the energy without vibration.

There are two relevant normal modes that are Coriolis-coupled, 
and, because the curvature of the
vibrational potential obtained by diagonalizing the Hessian matrix 
around 
the saddle point is small, one of the solutions of the coupled 
equation of vibrational motion becomes unstable.  
However, this way (i.e., use of the Hessian matrix) of evaluating the 
potential is only valid in the immediate vicinity of the 
saddle point, while the wavefunction has a much larger 
extension, over which we should approximate the vibrational potential 
(inset in Fig. \ref{fig:dist6} a) by a parabola.  
We can then fit the curvature of the vibrational potential in
the region of about 20\% of the mean radius, which 
is a few percent larger than the 
curvature obtained by 
the Hessian matrix, and indeed makes the vibrational motion stable.  
This is how we find the ground state energy in the six-fold case.  

For the five-fold configuration with a center electron, 
we have obtained the ground-state
energy in the same way as for $N=3, 4$.  We compare the ground-
state energies of six-fold and five-fold
configurations at each value of $L$.  
In  Fig.~\ref{fig:dist6} we have plotted the lowest of the two possible 
energies.  The result agrees very well with 
the exact diagonalization and
the difference between the ground-state energies 
is 0.4\% or less around $L=60$. 

\section{Conclusion and discussions}
In this paper, we have confirmed the accuracy of the electron-molecule 
description of interacting electrons in a quantum dot in high
magnetic fields for four, five and six electron systems.  
For four spin unpolarized electrons, there are two sequences of magic numbers
that correspond to two RVB-like spin configurations.
For five electrons, accurate ground state
energies are obtained by considering both $N$-fold and $(N-1)$-fold 
molecules.  
For six electrons, we have shown that the quantum
mixing between states localized on classically 
distorted configurations 
leads to the symmetry expected from the exact diagonalization
results. 


\end{document}